\documentclass[twocolumn,aps]{revtex4}
\usepackage{amsfonts,epsfig}
\def\g5{\gamma^5}

\def\d4k{{d^4k\over (2\pi)^4}}

\newcommand{\beq}{\begin{eqnarray}}
\newcommand{\eeq}{\end{eqnarray}}
\newcommand{\beqno}{\begin{eqnarray*}}
\newcommand{\eeqno}{\end{eqnarray*}}
\def\lsim{\mathrel{\rlap{\lower4pt\hbox{\hskip1pt$\sim$}}
    \raise1pt\hbox{$<$}}}         
\def\gsim{\mathrel{\rlap{\lower4pt\hbox{\hskip1pt$\sim$}}
    \raise1pt\hbox{$>$}}}         

\begin{document}
%
\title{Single and Double Peripheral Production of Sigmas in Proton Proton 
Collisions}
\author{Leonard S. Kisslinger\\
        Department of Physics,\\ \vspace{3mm}
       Carnegie Mellon University, Pittsburgh, PA 15213\\
       Wei-hsing Ma and Pengnian Shen\\
CCAST(World Laboratory) and Institute of High Energy Physics, Academia Sinica\\
               Beijing, P. R. China}
\indent
\begin{abstract}
The Pomeron, which dominates high energy elastic and diffractive  
hadronic processes, must be largely gluonic in nature. We use a recent
picture of a scalar glueball/sigma system with coupling of the sigma
to glue determined from experiment to predict strong peripheral sigma
production in the pp$\pi^0\pi^0$ and double sigma production in the
pp$\pi^0\pi^0\pi^0\pi^0$ channels.
\end{abstract}
\maketitle
\vspace{5mm}

\noindent
PACS Indices: 12.40.Gg, 12.38.Lg, 13.85.Dz, 13.60.Le
%
%
\section{Introduction}
\hspace{.5cm}

  The Regge formalism is successful for treating high energy elastic 
scattering and diffractive hadronic processes, and  all known hadrons 
seem to lie on Regge trajectories, however, it has long been known that 
the treatment of these high energy processes is not consistent with the 
Regge meson trajectories \cite{pcol}. Pomeron exchange
dominates these high energy processes, which means in terms 
of Quantum Chromodynamics (QCD) that the Pomeron is not related to
meson-like $\bar{q}q$ states, but is gluonic in nature. There has been 
a great deal of work in recent years in studying the gluonic structure
of the Pomeron. QCD perturbation theory has been used to derive the
hard or BFKL Pomeron (see Ref.\cite{lip97} used to treat high momentum
processes. The soft Pomeron, which we call the Pomeron, which can account
universally for elastic and peripheral process, must be treated by
nonperturbative QCD (see Ref.\cite{blt03} for a review of nonperturbative
QCD treatments, including the use of QCD instantons\cite{kkl01}).

   Peripheral processes, which correspond to the
production of low momentum particles from Regge trajectories, at high 
energies are given by the emission of the peripheral particles from the 
Pomeron, often referred to as Double Pomeron Exchange. 
Since the Pomeron is gluonic in nature, peripheral production 
at high energy is given by
the coupling  of the peripheral particles to the gluonic field.

   In the present work we study single and double sigma peripheral 
production in high energy proton proton (pp) collisions, using  an 
external field method to derive these processes.
It is based on the model\cite{lk1,lk2} that 
there exists a light scalar glueball strongly coupled to 
the I=0 two-pion system, which we call the glueball/sigma model,
and our recent work\cite{km} that this system might lie on the 
daughter trajectory of the Pomeron.  The sigma/glueball model is 
proposed based on three observations: 1) at low energies the
the scalar-isoscalar $\pi-\pi$ sysem is observed 
in $\pi-\pi$ scattering to be a Breit-Wigner resonance\cite{zb}, which we 
call the sigma; 2) the two-sigma channel is large in
scalar glueball decay\cite{bes}; and 3) in QCD sum rule calculations we
find\cite{lk1} a light scalar glueball far below the coupled scalar
glueball-meson systems, which we find correspond to the f$_0$(1370) and
f$_0$(1500). Our proposed glueball/sigma resonance is a coupled-channel
glueball-2$\pi$ system with a mass and
width both about 400 MeV. With this picture it was predicted\cite{lk2}
that there will be found a large branching ratio for the decay of the
P$_{11}$(1440) baryon resonance to a sigma and a nucleon.

  Three quantities are necessary for the calculation of the cross 
sections: 1) the Pomeron-Nucleon vertex, 2) the Pomeron propagator, 
and 3) the sigma-Pomeron vertex.
In our earlier work\cite{km} we showed that the coupling of the 
Pomeron to the nucleon can be predicted using the glueball/sigma model
with no free parameters, and that this coupling agrees within expected
errors with a phenomenological 
Pomeron exchange model\cite{dl} that is consistent with many high energy
experiments. We use this phenomenological model of the Pomeron-Nucleon
vertex. The sigma-gluonic coupling, obtained from the glueball/sigma 
picture of Ref \cite{lk2} gives the sigma-Pomeron vertex. Since in the
present work we give our results as the ratio cross sections of the 
peripheral production processes to the elastic scattering with similar
kinematics, many details of the Pomeron formalism are not important.
We also note that recent work suggests\cite{lm} that the $\xi(2230)$, 
if it turns out to be a tensor glueball, might lie on the Pomeron itself,
while in our formalism the tensor glueball lies on the trajectory of
the light scalar glueball\cite{km}, and should be found at about 2.8 GeV

   In Section II we briefly review the Pomeron formalism for elastic pp
scattering. In Sections III and IV we derive the cross section for single
and double sigma peripheral production, respectively. In Section V we
give our conclusions that there are large branching ratios that can be 
tested in experiment.
   
\section{Elastic pp Scattering Via Pomeron Exchange}
\hspace{.5cm}

  In this section we briefly review the formalism for elastic pp scattering
at high energy, which is known to be dominated by Pomeron exchange. The
two ingredients of the quite successful phenomenological model of Pomeron
exchange for high energy elastic nucleon-nucleon scattering\cite{dl} are the
Pomeron-Nucleon vertex, $V^{P-N}_\mu$,
\beq
\label{vpn}
       V^{P-N}_\mu & = & \beta \gamma_\mu F_1(t) \;\; ,
\eeq
and the Pomeron propagator, $D^P$. The constant $\beta$ is known to be
approximately 6 GeV$^2$, but will not be needed in the present work.      
As depicted in Fig. 1a, the form of the elastic proton-proton scattering 
amplitiude with Pomeron exchange is given by
\beq
     A^{pp} & = & V(t) D^P(t,s) V(t),
\label{1}
\eeq
where $t=(p_1 -p_1')^2$, $s=(p_1 +p_2)^2$, and V(t) is given by $\beta$ 
and $F_1(t)$ of Eq.(\ref{vpn}). We write the Pomeron propagator  D$^P$ 
using the notation of QCD sum rules that is used to study glueballs\cite{lk1}
\beq
    D^P(q,s) & = & \int d^4x e^{iq\cdot x}<0|T[[G(x)G(x)] \nonumber \\
              && [G(0)G(0)]]|0> \;\; ,
\label{2}
\eeq
where [G(x)G(x)] is a symbolic form for the current of the Pomeron.

\begin{figure}
\begin{center}
\epsfig{file=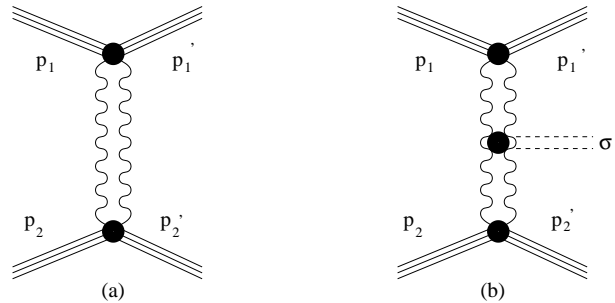,width=8cm}
\caption{a)Elastic p-p scattering with Pomeron exchange , b) peripheral 
production with double Pomeron exchange.}
{\label{Fig.1}}
\end{center}
\end{figure}

  This formalism for elastic scattering will be extended to peripheral
production using an effective field methon in the following two
sections.

\section{Single Sigma Peripheral Production in Proton Proton Collisions}
\hspace{.5cm}

In this section we derive the cross section for peripheral sigma production
using our glueball/sigma model.
With a notation similar to Eq(\ref{1}), the amplitude for the peripheral
production of $\sigma \rightarrow \pi\pi$ as shown in Fig. 1b is given by
\beq
     A^{pp\sigma} & \simeq & V(t_1) \bar{D}^P_\sigma(t_1,t_2,s) V(t_2),
\label{3}
\eeq
where $\bar{D}^P_\sigma$ is the propagator of the exchanged Pomeron coupled 
to a propagating $\sigma$ (which decays to the I=0 2$\pi$ resonance), which
is often called a double Pomeron, $t_1=(p_1 -p_1')^2$ and $t_2=(p_2 -p_2')^2$. 
Although $\bar{D}^P_\sigma$ depends 
on the momentum transfers to the two interacting nucleons, $t_{1}$ and 
$t_{2}$, which are different from the momentum transfer, t, for elastic  
scattering with the same $s$, the sigma meson for peripheral production 
carries a very small momentum $p_{\sigma}$, so that $t_{1} \simeq t_{2} 
 \simeq t$ unless  $t_{1},t_{2} \leq M_\sigma^2$. As an example,
for $p_\sigma$ = 0 and $M_\sigma$ =
400 MeV, in the center of mass system $t_1 = t_2 +$ 0.16 GeV$^2$.
For this reason we use the approximation that 
\beq
\label{dp}
   \bar{D}^P_\sigma(t_1,t_2,s) &\simeq & \bar{D}^P_\sigma(\bar{t},p_\sigma,s)
 \; ,
\eeq
with $\bar{t} = (t_1+t_2)/2$ and $p_\sigma$ is the momentum of the  $\sigma$ 
Even for modest energies and small momentum transfers, such as the 1985 
CERN experiment looking for glueballs in pp collisions at 
$\sqrt{s}$ = 63 GeV\cite{cern}, one could use an 
average value, as in Eq.(\ref{dp}),with only small changes in our results.
Note that our calculations are for 50 GeV protons and the sigma momentum is 
of the order of 0.3 GeV, and that in our work we use low-energy theorems, so 
that only very low energy $\sigma$ production can be treated in our model.
This also allows the use of the external field method, which is described
next.

Using the external field
method with the sigma treated as an external field we write the Pomeron
propagator coupled to a $\sigma$ as
\beq
 \bar{D}^P_\sigma(q,p_\sigma,s) & = & \int d^4x 
e^{iq\cdot x}<0|T[[G(x)G(x)] \nonumber \\
     && [G(0)G(0)]]_\sigma|0> \;\; .  
\eeq
Assuming factorization, we use $ [[G(x)G(x)][G(0)G(0)]]_\sigma \simeq
[G(x)G(0)][G(x)G(0)]_\sigma$, with $<T[G(x)G(0)]>$ the gluon 
propagator and $<T[G(x)G(0)]_\sigma>$ an external field
expression for the sigma coupled to the gluon, giving
\beq
\label{4}
 \bar{D}^P_\sigma(q,p_\sigma,s) & = & \int d^4k D^P(k,s) g_\sigma 
F_\sigma(q-k,p_\sigma),
\eeq
with $F_\sigma(q-k,p_\sigma)$ the sigma-gluon vertex form factor and $g_\sigma$
the coupling constant of the sigma to the gluon. The value of 
$g_\sigma$ was extracted from experiment in Ref.\cite{lk2}
from the glueball/sigma model, with its latest version
including instanton effects\cite{kj} in essential agreement with the earlier
model. The method is to use the external field method 
for the sigma current coupled to the gluon propagator\cite{lk2}:
\beq
  <G^a_{\alpha\beta} J_\sigma G^{\alpha\beta}_a> & = &
  g_\sigma <G^a_{\alpha\beta} G^{\alpha\beta}_a> \nonumber
\eeq
Assuming that the sigma $\pi \pi$ resonance arises from light glueball
pole, $g_\sigma$ is given by the width of the pole, 
or $g_\sigma \simeq$ 400 MeV.
For low-momentum sigmas, which we are assuming, $F_\sigma(q-k) \simeq
 \delta^4(q-k)$. Including propagation of the $\sigma$ to the $\pi\pi$
vertex we obtain for the Pomeron propagator with a produced $\sigma$
with momentum p$_\sigma$ :
\beq
 \bar{D}^P_\sigma(\bar{t},p_\sigma,s) & \simeq & g_\sigma G_\sigma(p_\sigma) 
D^P(\bar{t},s).
\label{5}
\eeq
In Eq.(\ref{5}) G$_\sigma$ is the Breit-Wigner resonance propagator
of the sigma. Introducing the appropriate phase space factors we find 
from Eqs.(\ref{1}, \ref{3}, and \ref{5}) the relationship
\beq
      \frac{d^2 \sigma^{pp\sigma}/dt dE_\sigma}{d\sigma^{pp}/dt}
 & = & \frac{1}{4 \pi^2}
\frac{g_\sigma^2 p_\sigma}{(p_\sigma^4 + M_\sigma^2 \Gamma_\sigma^2)}
 {\cal F}(t_1,t_2,\bar{t}) \; ,
\label{6}
\eeq
with
\beq
\label{F}
         {\cal F}(t_1,t_2,\bar{t}) &=& \frac{F^2(t_1) F^2(t_2)}{F^4(\bar{t})}
\; ,
\eeq
where $t_1 = t_2 +p_\sigma^2 + 2p_\sigma \cdot (p_2 - p_2')$, and
we have included the fact that both gluons can radiate sigmas.
The energy/momentum structure of Eq.(\ref{6}) can be understood by considering
the dispersion relation for $D^P_\sigma$, and recognizing that in our
approximations we assume that it is dominated by the pole and that at low 
momentum for the $\sigma$ the numerator of the pole is independent of energy.
The function $F(t)$, which gives the $t$-dependence of the Pomeron-quark
vertex, is taken from the phenomenological fits of Ref.\cite{dl}:
\beq
\label{fdl}
     F(t) &=& \frac{4 M_N^2 -2.79 t}{4 M_N^2 - t}\frac{1}{(1-t/.71)^2} \; .
\eeq
The results for the ratio of cross sections in the center of mass frame are 
shown in Fig. 2.
\begin{figure}
\begin{center}
\epsfig{file=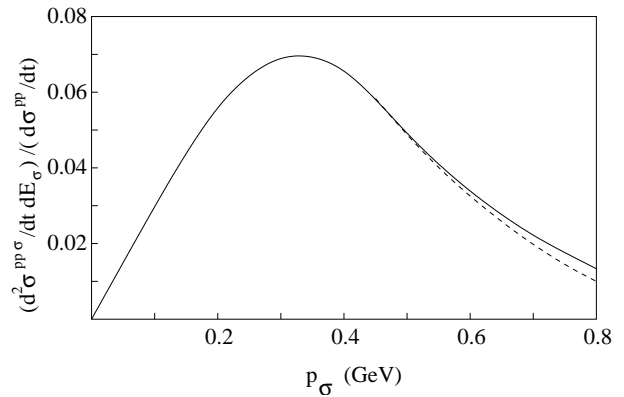,width=8cm}
\caption{Ratio of differential cross sections of $pp \rightarrow pp\sigma$
process to elastic $pp \rightarrow pp$ process. Dashed line shows the Ratio
without the factor ${\cal F}$, Eq.(\ref{F}).}
{\label{Fig.2}}
\end{center}
\end{figure}
As shown by comparing the dashed curve to the solid curve in Fig. 2, for 
values of the sigma momentum consistent with
peripheral production, the effects of $t_1 \neq t_2$ are very small.
For this reason we drop the factor $ {\cal F}(t_1,t_2,\bar{t})$ in the
rest of the calculations.

For comparison with experiment we rewrite our results in terms of
the $\sigma$ rapidity distribution which is a function
of the transverse momentum of the $\sigma$. With the standard definition
the rapidity of the $\sigma$ and the $\sigma$ energy, they are given by 
\beq
\label{7}
           y & = & tanh^{-1}(\frac{p_{\sigma z}}{E_\sigma}), \\
          E_\sigma & = & \sqrt{m_\sigma^2 + p_{\sigma\perp}^2}cosh(y),
\eeq
where $p_{\sigma\perp}  = \sqrt{p_x^2 + p_y^2}$ is the transverse 
momentum of the $\sigma$.
Using the approximation that the t variable is the same as for elastic
p-p scattering for peripheral production, as explained above, we integrate 
over the t variable to obtain from Eq.(\ref{6})
\beq
\label{8}
        \frac{d \sigma^{pp\sigma}}{dy} & = & \sigma^{pp}_{tot}
 \frac{g_\sigma^2}{4\pi^2}sinh(y) \\
      &&\frac{ \sqrt{(m_\sigma^2 + p_{\sigma\perp}^2)
 ((m_\sigma^2 + p_{\sigma\perp}^2)cosh^2y-m_\sigma^2)}}
 {((m_\sigma^2 + p_{\sigma\perp}^2)cosh^2y-2m_\sigma^2)^2 +m_\sigma^2
 \Gamma_\sigma^2} \nonumber \; \; .
\eeq
Using the experimental fit to the total elastic p-p cross section at high
energy\cite{tot}, $\sigma^{pp}_{tot} = 21.70 s^{0.0808} + 56.08 s^{-0.4525}$
mb, which is similar to the fit obtained with the Pomeron/Reggeon model of 
Ref. \cite{dl}, 
we obtain the results shown in Fig. 3.        
\begin{figure}
\begin{center}
\epsfig{file=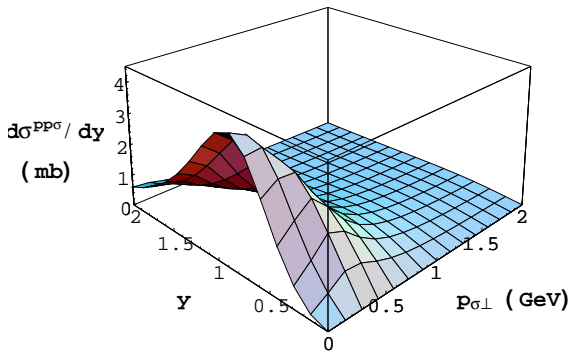,width=8cm}
\caption{Rapidity and transverse momentum distribution for sigma production.}
{\label{Fig.3}}
\end{center}
\end{figure}
As seen in Fig. 3., in our model the cross section for peripheral production 
near y = 1.0 and low momentum transfer has a peak of about 4.0 mb.
Note that the charged $\pi^+\pi^-$ channel 
might not be as satisfactory because of the $\rho$-meson background from 
processes at the nucleon vertices. Even though $\rho$-mesons cannot be 
produced in Pomeron diffractive production, they can be produced by the
scattered nucleons at a Pomeron-nucleon vertex, which makes the 
interpretation of experiment more difficult in some reactions.

Recently there have been extensive experiments by the WA102 collaboration 
in which centrally produced $\pi^+ \pi^-$ and $\pi^0 \pi^0$ systems have 
been analyzed\cite{wa}.
Although the publications have not discussed the detection of what one now 
calls $\sigma$, one can see structure in the low energy $\pi\pi$ systems
that resemble the recent charm D$^+$ decay experiments of the E791 
collaboration\cite{E791} and the $\tau^-$ decay of the CLEO 
collaboration\cite{cleo}, where the broad 400 MeV $\sigma$ was seen. 
It is also interesting to note that a Regge pole analysis\cite{mp} is in
agreement with the low-energy behavior of the CERN AFS 
experiment\cite{cern}. Althought the physics of that fit to the data is
quite different from our glueball/sigma picture, the parameterization 
of  the $\pi \pi$ amplitude is similar to ours and our model should also
be consistent with the data if the branching ratio of the low-energy
$\pi \pi$ channel were extracted. Also, very recently the sigma, with
the mass and width parameters consistent with our glueball/sigma model
has been observed\cite{bes04} in the $J/\Psi\rightarrow \omega \pi^+ \pi^-$
decay

\section{Double Peripheral Production of Sigmas in Proton Proton Collisions}
\hspace{.5cm}

In this section we study the high energy reaction $pp \rightarrow pp\sigma
\sigma$ with low-momentum peripheral production via Pomeron exchange. Using
the same methods as in the previous section,
the amplitude for this peripheral double sigma production can be written as
\beq
     A^{pp\sigma_1\sigma_2}(t_1,t_2,s) & = & 
V(t_1) \bar{D}^P_{\sigma_1\sigma_2}(t_1,t_2,s) V(t_2),
\label{2d}
\eeq
where $\bar{D}^P_{\sigma_1\sigma_2}$ is the propagator of the exchanged 
Pomeron coupled to two $\sigma$s, $t_1=(p_1 -p_1')^2$, 
$t_2=(p_2 -p_2')^2$, and $p_{\sigma_1},p_{\sigma_2}$ are the momenta
of the two sigmas. Note that our calculations are 
for 50 GeV protons and the sigma momenta are of the order of 0.3 GeV
for sigma peripheral production, so that the momentum transfer at each
nucleon vertex is approximately t, the momentum transfer for elastic 
pp scattering. Taking advantage of the very low momentum of the sigmas,
we use the effective field method to estimate the Pomeron propagator
coupled to two sigmas. 
\begin{figure}
\begin{center}
\epsfig{file=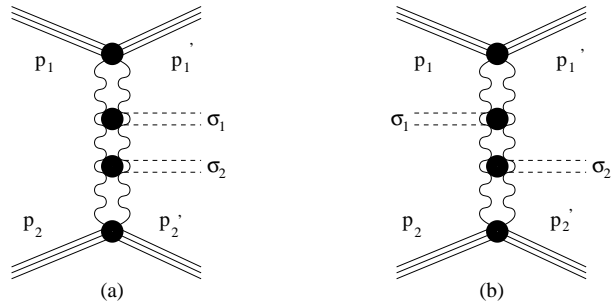,width=8cm}
\caption{a)Peripheral production from each gluonic leg of pomeron; (b)
Double peripheral production from one leg}
{\label{Fig.4}}
\end{center}
\end{figure}

  Corresponding to Fig. 4b, the Pomeron propagator with a sigma coupled to 
each gluonic leg is
\beq
 \bar{D}^P_{\sigma_1\sigma_2}(q,s) & = & \int d^4x 
e^{iq\cdot x}<0|T[[G(x)G(0)]_{\sigma_1} \nonumber \\
                    && [G(x)G(0)]]_{\sigma_2}|0> \;\; . 
 \label{3d}
\eeq
Using factorization and neglecting the form factor for the sigma-gluonic
coupling as part of our low-momentum calculaiton, we find
\beq
\label{4d}
\bar{D}^P_{\sigma_1\sigma_2}(\bar{t},s) & = &
 g_{\sigma}^2 G_\sigma (p_{\sigma_1})G_\sigma(p_{\sigma_2})D^P(\bar{t},s)\; ,
\eeq
with $\bar{t} = (t_1+t_2)/2$ and $p_\sigma$ is the momentum of a $\sigma$,
and the Pomeron propagator is evaluated at $\bar{t}\simeq t_1\simeq t_2$
(as we discuss below). g$_\sigma$ is the $\sigma$-gluon coupling constant 
derived in Ref \cite{lk2} and G$_\sigma$ is the Breit-Wigner resonance 
propagator of the sigma. Using the factorization approximation it is easy 
to see that the process with both sigmas coming from the same gluonic leg, 
pictured in diagram of Fig. 4a gives an equal contribution. There is
also an equal contribution from the process with the two sigmas coupled
to the other leg. This gives us for the amplitude of the double-sigma
peripheral production
\beq
\label{5d}
      A^{pp\sigma\sigma}(t_1,t_2,s) & = & 3
g_\sigma^2 G_\sigma(p_{\sigma_1})G_\sigma(p_{\sigma_2})A^{pp}(\bar{t},s)\; 
\eeq
in terms of A$^{pp}$, the pomeron amplitude for pp elastic scattering.
Introducing the appropriate phase space factors we find for the ratio
of the double peripheral sigma production to the pp elastic scattering
\beq
\label{6d}
      \frac{\frac{d^3\sigma^{pp\sigma\sigma}}{dt dE_{\sigma_1} dE_{\sigma_2}}}
{\frac{d\sigma^{pp}}{dt}}
 & = &
\frac{9 g_\sigma^2}{4  \pi^2} \frac{p_{\sigma_1}}
{(p_{\sigma_1}^4 + M^2_{\sigma_1} \Gamma^2_{\sigma_1})} \\
             && \frac{p_{\sigma_2}}
{(p_{\sigma_2}^4 + M^2_{\sigma_2} \Gamma^2_{\sigma_2})}
{\cal F}(t_1,t_2,\bar{t}) \nonumber \; ,
\eeq
with  ${\cal F}(t_1,t_2,\bar{t})$ given in the previous section. For 
$p_\sigma$ limited to 0.3 GeV to be consistent with peripheral production, 
the factor ${\cal F}$ is close to unity, as shown in Fig. 5.  
\begin{figure}
\begin{center}
\epsfig{file=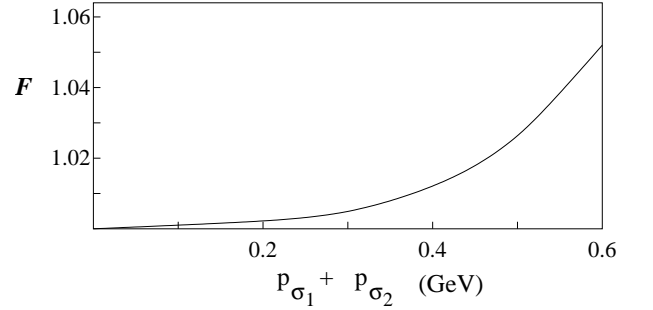,width=8cm}
\caption{The Factor ${\cal F}$ of Eq.(\ref{6}) for peripheral production.}
{\label{Fig.5}}
\end{center}
\end{figure}
As one can see the factor is close to unity, and we drop in the rest 
of this paper.

In Fig. 6 we show the ratio defined in Eq.({\ref{6d}) in terms of the
transverse momenta of the two sigmas. 
\begin{figure}
\begin{center}
\epsfig{file=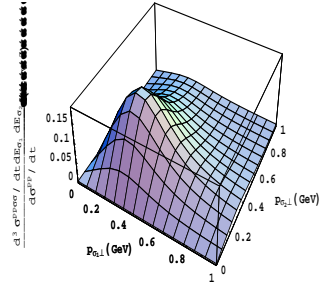,height=4cm,width=8cm}
\caption{Ratio of differential cross sections of $pp \rightarrow 
 pp\sigma_1 \sigma_2$ process to elastic $pp \rightarrow pp$ process.}
{\label{Fig.6}}
\end{center}
\end{figure}

For comparison with experiment we rewrite our results in terms of
the rapidity distributions of the two $\sigma$s.
Using the approximation that the t 
variable is the same as for elastic pp scattering for peripheral 
production, and integrating over the t variable one obtains 
from Eq.(\ref{6d})
\beq
\label{7d}
        \frac{d^2 \sigma^{pp\sigma\sigma}}{dy_1dy_2} & = & \sigma^{pp}_{tot}
 \frac{9 g_\sigma^4}{(2\pi)^4}sinh(y_1)sinh(y_2) \nonumber  \\
             && \frac{\sqrt{(m_\sigma^2 + p_{\sigma_1 \perp})E_{\sigma_1}^2 
-m_\sigma^2)}}
 {(E_{\sigma_1}-m_\sigma^2)^2 +m_\sigma^2
 \Gamma_\sigma^2}  \\
     & & \frac{\sqrt{(m_\sigma^2 + p_{\sigma_2 \perp}^2)E_{\sigma_2}^2
  -m_\sigma^2)}}
 {(E_{\sigma_2}^2 -m_\sigma^2)^2 +m_\sigma^2
 \Gamma_\sigma^2} \nonumber \;\; ,
\eeq
with the standard definition of rapidity, $ y  = tanh^{-1}(p_{\sigma z}/
E_\sigma),  E_\sigma = \sqrt{m_\sigma^2 + p_{\sigma\perp}^2}cosh(y)$. 

For our numerical plots shown in figures 7 and 8 we also use the experimental 
fit to the total elastic p-p cross section at high
energy\cite{tot}, $\sigma^{pp}_{tot} = 21.70 s^{0.0808} + 56.08 s^{-0.4525}$
mb.

In Fig. 7 The double rapidity distribution for the double peripheral sigma
production is shown with the transverse momentum of each sigma 
approximately 0.3 GeV.

In Fig. 8 the differential cross section with respect 
to the first sigma's rapidity is shown for important values of its 
transverse momentum, with the second sigma having transverse 
momentum approximately 0.3 GeV and rapidity 0.7. The plot is obtained from
Eq.(\ref{7d}) with y$_2$ fixed.

\begin{figure}
\begin{center}
\epsfig{file=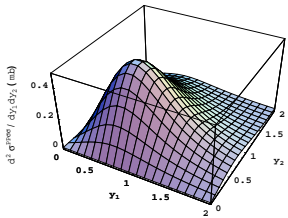,height=3cm,width=8cm}
\caption{Double differential cross sections of the
$pp \rightarrow pp\sigma\sigma$ process for $p_{ \sigma_1 \perp} \simeq 
p_{ \sigma_2 \perp} \simeq $ 0.3 GeV in terms of the y$_1$ and y$_2$ 
rapidities.}.
{\label{Fig.7}}
\end{center}
\end{figure}
\begin{figure}
\begin{center}
\epsfig{file=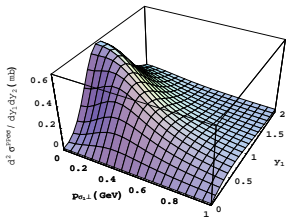,height=3cm,width=8cm}
\caption{Differential cross section with respect to y$_1$ of the
$pp \rightarrow pp\sigma\sigma$ process for $p_{\sigma_2 \perp} \simeq$ 
0.3 GeV and y$_2$ = 0.7 for important values of $p_{\sigma_1 \perp}$}
{\label{Fig.8}}
\end{center}
\end{figure}

It is seen that there is a large branching ratio for the double peripheral
production of the sigma $\pi\pi$ resonance. The experimental observation
of this process would be very valuable for improving our understanding
for gluonic interactions and the nature of the pomeron.

\section{Conclusions}
\hspace{.5cm}

We have derived the cross section for diffractive single and
double sigma production in
high energy proton-proton collisions. By mapping out the low-energy
spectrum of the two $\pi^0$s one will find the sigma resonance if this
glueball/sigma model is correct. If so, the production of sigmas can be
used as a signal for glueballs and hybrids as well as the Pomeron. 
There also would be important implications for sigma effects in charm
quark systems, where gluonic effects are known to be important.

The work was supported in part by the U.S. National Science Foundation
grants PHY-00070888, INT-9514190 and National Science Foundation of
China grants 10247004,10147202,90103020 as well as CAS Knowledge Innovation
Project No. KJCX2-N11.


\begin{references}
\bibitem{pcol}P.D.B. Collins, ``Introducton to Regge Theory (Cambridge
University Press, Cambridge, 1977), reviews Regge phenomenology and the 
Pomeron.
\bibitem{lip97}L.N. Lipatov, Phys. Rep. {\bf C286}, 132 (1997).
\bibitem{blt03}S. Bondarvenko, E. Levin and C-I. Tan, hep-ph/030623, to be
published in Nucl. Phys. {\bf A}.
\bibitem{kkl01}D. Kharzeev, Y. Kovcheyov and E. Levin, Nucl. Phys. 
{\bf A690}, 621 (2001).
\bibitem{lk1}L.S. Kisslinger, J. Gardner and C. Vanderstraeten, Phys. Lett.
{\bf 410}, 1 (1997).
\bibitem{lk2}L..S. Kisslinger and Z. Li, Phys. Lett. {\bf B445}, 271 (1999).
\bibitem{km}L.S. Kisslinger and W-H. Ma,  Phys. Lett. {\bf B485}, 367 (2000).
\bibitem{zb}B.S. Zou and D.V. Bugg, Phys. Rev. {\bf D50}, 591. (1995).
\bibitem{bes}BES Collaboration, J.Z. Bai et. al., Phys. Rev. Lett. {\bf 76},
3502 (1996).
\bibitem{dl}A. Donnachie and P. V. Landshoff, Nucl. Phys {\bf B231},189
 (1984); {\bf B244}, 322 (1984); Phys. Lett. {\bf B185}, 403 (1987).
\bibitem{lm}L.C. Liu and W-h. Ma, J. Phys. G: Nucl. Part. Phys. {\bf 26},
L59 (2000).
\bibitem{cern}T. Akesson et. al. Nucl. Phys. {\bf B264}, 154 (1986).
\bibitem{kj}L.S. Kisslinger and M.B. Johnson, Phys. Lett. {\bf B523}, 127
 (2001). 
\bibitem{tot}A. Donnachie and P. V. Landshoff, Nucl. Phys. {\bf B296}, 227
 (1992); J. R. Forshaw and D. A. Ross, ``Quantum Chromodynamics and the 
Pomeron'' (Cambridge University Press, Cambridge, 1997), p.9 
\bibitem{wa}WA102 Collaboration, D. Barberis et. al., Phys. Lett. {\bf B453},
 316; 325 (1999); {\bf B462}, 462 (1999); {\bf B467}, 296 (1999); {\bf B471}, 
 435; 440 (2000). 
\bibitem{E791}E791 Collaboration, E.M. Aitala et. al., Phys. Rev. Lett.
{\bf 86}, 770 (2001).
\bibitem{cleo}CLEO Collaboration, D. Asner et. al., Phys. Rev. {\bf D61} 
 012002 (2000).
\bibitem{mp}D. Morgan and M.R. Pennington, Phys. Lett. {\bf B137}, 411 (1984). 
\bibitem{bes04} BES Collaboration, M. Ablikim et. al., hep-ex/0406038, 
hep-ex/0404016.

\end{references}
\end{document}